\documentclass[manuscript,screen,sigconf, nonacm]{acmart}

\AtBeginDocument{%
  }

 \usepackage{multirow}
\usepackage{makecell}
\usepackage{longtable}
\usepackage{enumitem}
\usepackage{xcolor}
\usepackage{soul}
\usepackage{listings}
\usepackage{makecell}
\usepackage{float}
\usepackage{booktabs}
\usepackage{bbding}
\usepackage{makecell}
\usepackage{subcaption}
\usepackage{hyperref}
\usepackage{float}   
\usepackage{xparse}
\usepackage{tabularx}
\usepackage{subcaption}
\newcolumntype{R}{>{\raggedright\arraybackslash}X}

\AtBeginDocument{%
  }
\usepackage{csquotes}

\setcopyright{acmlicensed}
\copyrightyear{2018}
\acmYear{2018}
\acmDOI{XXXXXXX.XXXXXXX}
\acmConference[Conference acronym 'XX]{Make sure to enter the correct
  conference title from your rights confirmation email}{June 03--05,
  2018}{Woodstock, NY}
\acmISBN{978-1-4503-XXXX-X/2018/06}

\newcommand\wordcount{
    \immediate\write18{texcount -sub=section \jobname.tex  | grep "Section" | sed -e 's/+.*//' | sed -n \thesection p > 'count.txt'}
(\input{count.txt}words)}

\begin{document}

\title{Novice Reliance Calibration in AI-Assisted Decision Making: The Role of Explanations and Self-Assessment}

\author{Eun Jeong Kang}
\email{eunkang@infosci.cornell.edu}
\affiliation{%
  \institution{Cornell University}
  \city{Ithaca}
  \state{NY}
  \country{USA}
}

\author{Peter (Xianpi) Duan }
\email{duanx14@mcmaster.ca}
\affiliation{%
  \institution{McMaster University}
  \city{Hamilton}
  \state{Ontario}
  \country{Canada}
}

\author{Swati Mishra}
\email{swati@mcmaster.ca}
\affiliation{%
  \institution{McMaster University}
  \city{Hamilton}
  \state{Ontario}
  \country{Canada}
}

\renewcommand{\shortauthors}{Kang et al.}

\begin{abstract}
Artificial Intelligence (AI) tools are widely used to support decision making in tasks and domains where no immediate performance feedback is available. In these settings, users cannot learn to adjust their reliance behavior over time through trial and error. However, little is known about how novice users calibrate reliance on AI when external feedback is unavailable, or whether AI explanations can support calibration in its absence. We introduce reliance calibration as an organizing construct for studying how novice users dynamically adjust reliance behavior, and examine how AI explanations and meta-cognitive self-assessment shape it. Through a between-subjects study with 110 participants completing a clinical entity extraction task with AI assistance and limited performance feedback, we observe that novice users exhibit systematic drift toward over-reliance in the presence of explanations, while higher self-reported task understanding is associated with more selective reliance behavior. These results extend reliance calibration research into human-AI collaboration contexts without real-time performance signals and present actionable guidelines on designing AI tools that must support appropriate reliance in these settings. 
\end{abstract}



\maketitle





\section{Introduction}

A central challenge to the adoption of Artificial Intelligence (AI) systems in complex decision making tasks is achieving appropriate human reliance on AI.
For a productive human-AI collaboration, users of these systems must continue to accept accurate AI recommendations and override incorrect ones, a behavior known as \textit{appropriate reliance}\cite{schemmer_appropriate_2023, lee_trust_2004,bansal2019beyond}.
However, appropriate reliance is not a fixed state that can be characterized by a single decision on a single recommendation. 
As people interact with an AI system over repeated tasks, they continuously update how much they defer to or deviate from its suggestions~\cite{kahr2024understanding, srinivasan2026adjust}.
A user who initially accepts most AI recommendations may begin to reject more of them after encountering patterns that suggest errors.
Conversely, a user who starts skeptically may gradually defer to the system as its suggestions prove reliable.
We define this process as \textit{reliance calibration -- the degree to which a user's acceptance or rejection of AI suggestions aligns with the actual correctness of those suggestions over the course of an interaction session}. 
A well-calibrated user \textit{frequently} accepts suggestions when the AI is correct and rejects them when it is not.
A miscalibrated user, on the other hand, \textit{frequently} over- or under-relies for their decision making.

Prior research has examined how users with varying levels of \textit{task expertise} learn to calibrate their reliance over time\cite{ turner_calibrating_nodate, kahr2024understanding, tatasciore2025calibrating, srinivasan2026adjust}, and how design interventions such as \textit{AI explanations} influence this calibration behavior~\cite{bussone2015role,gunning2019xai,zhang2020effect}.
These studies, however, primarily focus on the setting where users receive \textit{explicit performance feedback}, defined by task-level \textit{signals that inform users whether their judgment (or the AI's suggestion) was correct after each decision}\cite{srinivasan2026adjust,nourani_role_2020,buccinca2024contrastive}.
In such settings, both performance feedback and explanations serve as learning signals that enable users to update their mental model of AI reliability and adjust subsequent actions accordingly. 

However, in most real-world deployments where AI is useful, it is challenging (and even redundant) to provide users with tailored signals about the correctness of their reliance decisions during or immediately after task completion. 
For instance, a clinician using an AI diagnostic tool, a quality control inspector using AI to detect surface defects in steel sheets, or an analyst labeling medical text to categorize them, may complete an entire interaction session without knowing whether their acceptance or rejection of AI suggestions was appropriate.
In such no-feedback environments, users draw on internal meta-cognitive resources to guide reliance decisions \cite{flavell1979metacognition, gajos2022people,ackerman2017meta}, reflecting on what they know and how confident they are in that knowledge, to determine whether to accept or override an AI suggestion.
Explanations also may not uniformly improve reliance calibration in these settings, as their effectiveness depends on the meta-cognitive resources users bring to the task \cite{vasconcelos2023explanations,anectodal}.
For instance,  expert users may use their task understanding and confidence as an internal calibration signal and verify their reasoning using explanations even when external feedback is absent \cite{spitzer2022training}.
However, this mechanism breaks down among \textit{novice users, whose domain expertise is insufficient to independently verify AI outputs and interpret explanations}, which is also where AI assistance is quite useful \cite{nourani_role_2020, dikmen2022effects}.

In this research, we investigate metacognitive mechanisms that drive reliance calibration in human-AI collaborative environments with novice users and no explicit performance feedback and the role of explanations in these settings.
We specifically focus on \textit{perceived competence} as the meta-cognitive factor and measure it as an aggregate of a user's perceived understanding of the task and their perceived confidence in their ability to perform the task\cite{paris2013metacognition}. 
This is important to study for novice users, who often hold miscalibrated and unstable perceptions about their own competence, and may be systematically misaligned with their actual ability to evaluate AI outputs \cite{kruger1999unskilled}.
We predict that novice users with higher perceived competence may engage more critically with AI suggestions and their explanations, while those with lower perceived competence may defer to the AI by default. 

Through a between-subjects study with 110 participants working on a clinical entity extraction task, we collected fine-grained interaction logs including acceptance, rejection, and editing of AI suggestions. 
Participants completed a sequence of annotation tasks with AI assistance and explanations but received no performance feedback during the session. 
We operationalize reliance calibration as a Brier-like miscalibration score \cite{glenn1950verification}, adapting this well-validated probabilistic scoring framework to measure the alignment between reliance decisions and AI correctness. 
We decompose this score into over- and under-reliance components, to support directional diagnosis of miscalibration in feedback-free environments.
Using these scores derived from participants' interaction logs, and pre-session and post-session self-report measures of perceived competence, we examine (1) how novice users' reliance calibration evolves across a session without performance feedback, (2) whether AI explanations improve calibration, and (3) how perceived competence predicts and moderates these effects.
Our results show that AI explanations were ineffective for improving reliance calibration when participants received no signals about the correctness of their current work. Instead, when participants perceived that they understood the clinical category entities, they demonstrated better reliance calibration.
This work contributes to the study of reliance calibration in no-feedback human-AI collaborative environments and provides a practical measurement framework and empirical evidence on the role of meta-cognitive factors such as perceived competence in reliance miscalibration among novice users.


\section{Related Work}
In this research, we build upon a growing body of literature across three primary areas. 
We first examine how reliance calibration has been conceptualized and measured in human-AI interaction. 
We then review studies on explanations and their impact on reliance behavior. 
Finally, we discuss the role of meta-cognition and self-assessment as individual-level predictors of how users engage with AI systems.

\subsection{Reliance Calibration in Human-AI Interaction}

Reliance is a behavioral construct captured through the actions users take in response to AI suggestions \cite{cao2022understanding, papenmeier2022s, parasuraman1997humans}, distinct from trust, which reflects a subjective psychological disposition rather than observable action \cite{van2020embedding, kim2023humans}.
Existing behavioral measures such as Agreement Score \cite{buccinca2020proxy, he2023knowing} and Switch Fraction \cite{lu2021human} have been used to study reliance across a range of factors including explanation usability \cite{nourani_role_2020}, cognitive load \cite{abdul2020cogam}, and prior beliefs \cite{nourani2021anchoring}. 
Studies on reliance calibration further examine how users optimize reliance in response to AI accuracy and accumulated experience, with the goal of achieving appropriate reliance \cite{yin2019understanding, zhang2020effect, kahr2024understanding, schemmer_appropriate_2023}.
However, existing research operates under three limiting assumptions. 
First, that users receive immediate correctness feedback after each decision; second, that users possess sufficient task knowledge to evaluate AI recommendations; and third, that reliance decisions are binary outcomes, overlooking partial reliance behaviors such as editing or augmenting AI suggestions. 
We address these gaps by proposing a session-level Reliance Calibration Score (RCS), adapted from the Brier Score \cite{glenn1950verification}, a validated probabilistic scoring framework that quantifies the mean squared difference between predicted probabilities and observed outcomes. 
Our adaptation treats reliance decisions as continuous inputs, accommodates partial reliance actions, and enables regression-based analysis of calibration trajectories without requiring real-time feedback. 
We further decompose RCS into over- and under-reliance components to support directional diagnosis of miscalibration.
We elaborate on this measure in Section~\ref{sec:measures}.

\subsection{AI Explanations and Their Effect on Reliance
}

Research in explainable AI has proposed various mechanisms for supporting reliance with some evidence suggesting that presenting the reasoning behind an AI recommendation can help users evaluate whether to accept or reject it~\cite{miller2019explanation, schoeffer2024explanations, anectodal}. 
Explanations have been shown to reduce uncertainty in interpreting AI recommendations and may be presented as highlighted text or data features \cite{robbemond2022understanding}, visual cues \cite{bansal2021does}, and concepts from task domain~\cite{mishra2021crowdsourcing}. 
However, the evidence that explanations improve reliance calibration is mixed~\cite{vasconcelos2023explanations,anectodal}. 
Several studies find that explanations do not meaningfully improve user performance and can even cause users to overlook their own judgment when AI output is present~\cite{nourani_role_2020, zhang2020effect, wang2021explanations, lu2024does}, while others demonstrate that when explanations are designed to promote engagement and active evaluation, they can reduce over-reliance~\cite{buccinca2021trust, buccinca2024contrastive}. 
A consistent finding across this literature is that the effect of explanations depends on whether users can meaningfully interpret them~\cite{kim_help_2023}. 
This condition is hardest to meet for novice users working on domain-specific tasks, who lack the expertise needed to evaluate explanation quality~\cite{fok2023search, langer2022look}. 
Additionally, in the absence of performance feedback, users cannot learn from errors to recalibrate how they use explanations over time, or rely on AI recommendations. 
Our study directly investigates this intersection by examining how explanations affect reliance calibration among novice annotators working without feedback on a complex task.

\subsection{Meta-Cognition and Self-Assessment in Human-AI Interaction}

Meta-cognition refers to an individual's ability to monitor and evaluate their own knowledge and performance, and has been widely studied as a determinant of decision-making quality \cite{flavell1979metacognition, dunning2011dunning}. 
In human-AI interaction, meta-cognitive factors such as self-assessment of domain knowledge and perceived competence shape reliance behaviors in distinct ways. 
Self-assessed domain knowledge influences whether users approach AI suggestions critically or deferentially, where users who believe they understand the subject matter are more likely to scrutinize AI output, while those with lower knowledge beliefs tend to defer \cite{chong2022human, he2023knowing}. 
Perceived competence, linked to task self-efficacy, governs willingness to exercise independent judgment, with higher competence beliefs associated with more active and critical engagement \cite{bandura1997self, chiang2022exploring}. 
For novice users, both factors are prone to miscalibration \cite{kruger1999unskilled, keren1991calibration}.
Overconfidence can lead to unwarranted rejection of correct AI suggestions, while underconfidence can produce uncritical acceptance of incorrect ones. Recent HCI research confirms that these factors moderate how users update reliance across interactions \cite{pataranutaporn2023influencing, li2025confidence, nourani_role_2020}, with self-confidence shown to shift in response to AI-expressed confidence rather than actual task performance \cite{li2025confidence}. In the absence of performance feedback, these meta-cognitive factors may serve as the primary regulators of reliance behavior, making them especially consequential in the feedback-free settings we study here.

\section{Research Questions}
In this study we investigate how novice users calibrate reliance on AI suggestions in a clinical annotation task where no immediate performance feedback is provided.
We specifically focus on how AI explanations and meta-cognitive factors such as perceived competence shape reliance calibration over time, through the following research questions:

\begin{itemize}
    \item \textbf{RQ1}: How does reliance calibration evolve across a task 
    session among novice users in the absence of performance feedback?
    \begin{itemize}
        \item \textbf{H1}: Reliance 
        calibration will deteriorate across the session, reflected in 
        increasing RC scores over time.
    \end{itemize}

    \item \textbf{RQ2}: Do AI explanations improve reliance calibration 
    relative to AI suggestions alone?
    \begin{itemize}
        \item \textbf{H2}: Users who receive explanations alongside AI 
        suggestions will demonstrate better reliance calibration, reflected 
        in lower RC scores, compared to those receiving suggestions without 
        explanations.
    \end{itemize}

    \item \textbf{RQ3}: In the absence of feedback, does users' perceived competence predict reliance calibration, and does it moderate the effect of AI explanations?
    \begin{itemize}
        \item \textbf{H3a}: Users with higher perceived competence will exhibit lower RC scores, reflecting more accurate alignment between 
        their reliance decisions and the correctness of AI suggestions.
        \item \textbf{H3b}: The calibrating effect of AI explanations will be moderated by perceived competence, with greater calibration expected in users with higher perceived competence.
    \end{itemize}

    \item \textbf{RQ4}: Within perceived competence, does self-reported 
    confidence in task ability predict better reliance calibration independently 
    of task understanding?
    \begin{itemize}
        \item \textbf{H4}: Self-reported task confidence will predict better 
        reliance calibration over and above task understanding, such that users who feel more capable of performing the task will show lower 
        RC scores even after controlling for their self-reported understanding.
    \end{itemize}
\end{itemize}

\section{Methodology}
 \label{method}
We selected a clinical entity extraction task that (1) requires specialized domain knowledge to accomplish accurately, (2) remains achievable with effort, stimulating participants' desire for challenge, and (3) is sufficiently complex to warrant AI recommendations.
To support this task we developed an interactive tool that provides AI suggestions alongside explanations for different entities across multiple clinical documents. 
We implemented a between-subjects experimental design, where participants completed a series of continuous entity extraction tasks, and manipulated the level of AI support across conditions. 
All study procedures were reviewed and approved by the Institution’s Research Ethics Board.

\begin{figure*}[t!]
\centering
  \includegraphics[width=\textwidth]{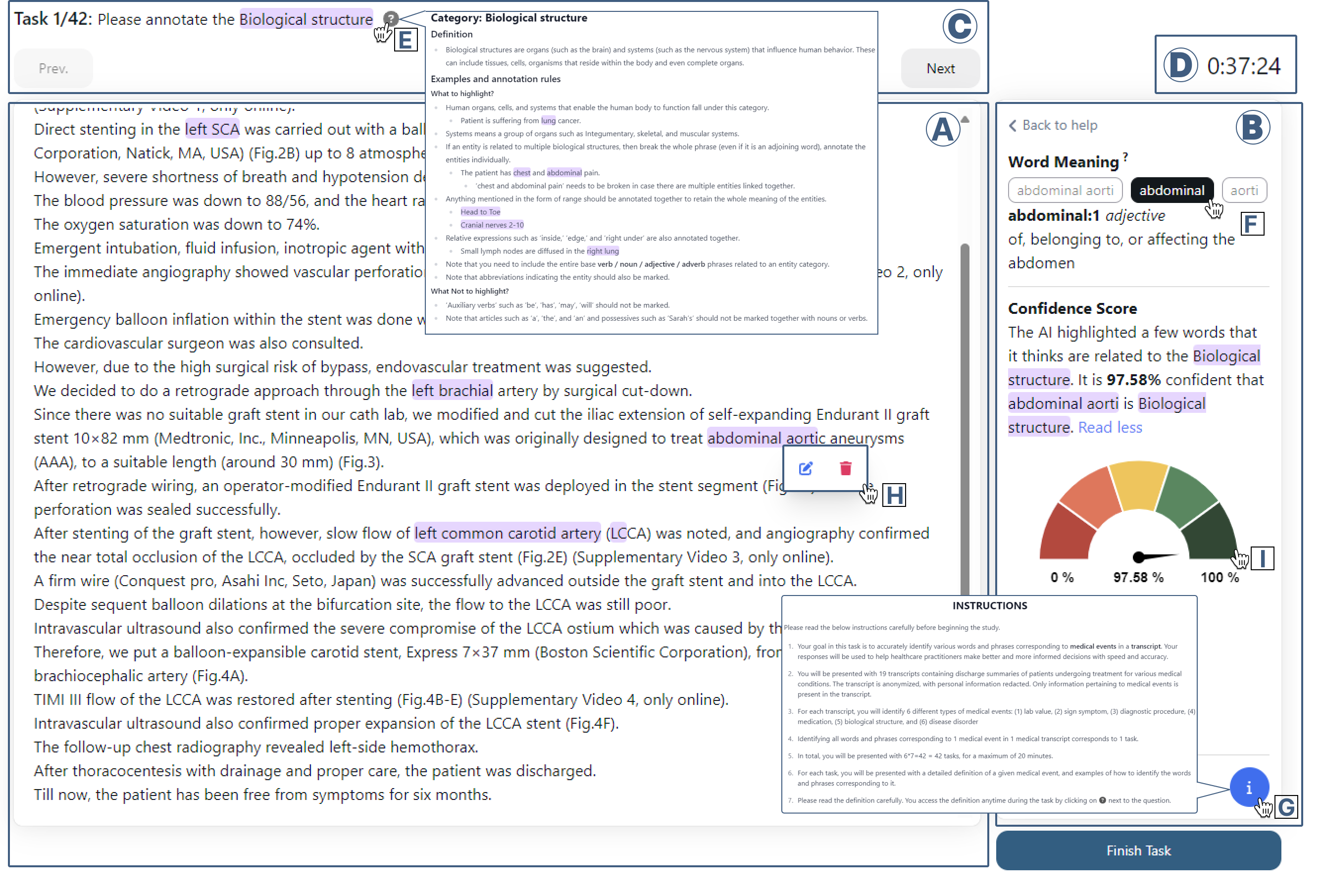}
  \caption{Interface to collaborate with the AI model; AI highlights words and phrases corresponding to a presented clinical category. In the \textit{No AI} condition, no highlights are presented. In the \textit{AI + EXP} condition, the AI confidence score is displayed when the user clicks on AI suggestions. In both the \textit{AI + EXP} and \textit{No AI + MD} conditions, a medical dictionary lookup tool (B) appears when the user clicks on words or AI suggestions. } 
  \label{fig:interface}
\end{figure*}

\subsection{Task}

Participants extracted clinical words and phrases corresponding to six different (6) categories (Signs \& Symptoms, Diagnostic Procedures, Lab Value, Medication, Disease Disorder, and Biological Structures) from anonymized patient discharge summary documents~\cite{Caufield2020}.
The task was to identify and highlight text spans that correspond to the clinical category, for instance, highlighting ``brachial artery'' for the ``biological structure'' category for a given document. 
For accurate entity extraction, the participants must first understand the clinical categories and then correctly identify and highlight terms from the summary.  

In conditions with AI support, the model's predictions were presented as highlighted text spans as recommendations. 
Participants could then accept, reject, or edit these suggestions, or add new highlights for terms they believed the AI had missed. 
To simulate real-world scenarios, the interface provided no immediate feedback, requiring participants to independently assess the AI's accuracy.
We used a BERT model fine-tuned for the Named Entity Recognition (NER) task on the MACCROBAT 2018 dataset comprising 200 documents annotated by clinical experts across 24 clinical categories~\cite{Caufield2020}.
Participants performed the task on 20 documents from the MACCROBAT 2020 dataset that were \textit{not} used in fine-tuning the BERT model.   

In the condition where AI explanations are provided, the AI explanation comprised (1) a Medical Dictionary \cite{mw2024}, that provided accurate definitions of all clinical terms in the document, and (2) Confidence Score~\cite{zhang2020effect, mishra2023human} that indicated the probability of a highlighted word belonging to a clinical category, as identified by the AI model.

\subsection{Interface}

We developed an interactive system for collaborating with the AI model on the entity extraction task (Figure~\ref{fig:interface}), with embedded task instructions (Figure~\ref{fig:interface}-(G)). 
The interface presents one document and clinical category at a time (Figure~\ref{fig:interface}-(A)) alongside any AI recommendations as highlighted text. 
Participants can accept AI-highlighted text, delete it or edit it via the Delete and Edit buttons (Figure~\ref{fig:interface}-(H)), or highlight additional words they believe to be missed. 
If newly added text overlaps with existing highlights, the system automatically merges the regions and retains the most recent selection.

To support task comprehension, the interface provides category definitions (Figure~\ref{fig:interface}-(E)) describing each clinical category with examples of qualifying and non-qualifying words and phrases, for instance, explaining what constitutes a ``biological structure'' in a medical context.
These definitions were curated from guidelines used in prior medical dataset curation studies~\cite{uzuner20112010, sun2013evaluating, nci2024, googleMed2020, funkhouser2020pathology}. 
The task instructions (Figure~\ref{fig:interface}-(G)) and category definitions (Figure~\ref{fig:interface}-(E)) windows opened automatically when each new category was first presented to the user, ensuring exposure to participants at least once, and remained accessible throughout the task. 
Participants navigated task sequences using Next and Prev buttons (Figure~\ref{fig:interface}-(C)) and submitted completed tasks via the Submit button.
The explanation dashboard included a \textit{confidence score} (CS) (Figure~\ref{fig:interface}-(I)) and a \textit{medical dictionary} (MD) (Figure~\ref{fig:interface}-(F)) to help participants look up definitions of unfamiliar words and interpret the CS scores~\cite{mishra2023human, mw2024}. 
The interface was built using Next.js and Flask, with all user interactions recorded in a MySQL database.

\subsection{Experimental Conditions}

\begin{table*}[t!]
\centering
\resizebox{0.98\textwidth}{!}{%

    \begin{tabular}{lcccc}
    \toprule
    & \makecell{Category Definition} & \makecell{AI Recommendation} & \makecell{AI Confidence Score}
    & Medical Dictionary \\
    \cmidrule(lr){2-2}\cmidrule(lr){3-3}
    \cmidrule(lr){4-4}\cmidrule(lr){5-5}
    No AI & \Checkmark & - & - & - \\
    No AI + MD & \Checkmark & - & - & \Checkmark \\
    AI & \Checkmark & \Checkmark & -& - \\
    AI + Explanation & \Checkmark & \Checkmark & \Checkmark & \Checkmark \\
    \bottomrule
        
    \end{tabular}
    }
    \vspace{0.3cm}
    \caption{Experimental conditions treated by AI and explanations. }
    \label{fig:enter-label}
\end{table*}


We employed four (4) between-subjects conditions to examine how AI recommendations and explanations independently affect reliance calibration.
These included (1) \textit{No AI} baseline, (2) \textit{No AI + Medical Dictionary} to isolate the effect of domain support without AI recommendations, (3) \textit{AI Only} providing recommendations without explanations, and (4) \textit{AI + Explanation} providing both recommendations and explanations. 
We did not include an AI + Confidence Score only condition, as interpreting confidence scores without the contextual support of a medical dictionary would place unreasonable cognitive demands on novice users and confound reliance effects with task difficulty. 
All conditions included task instructions and category definitions (see Table \ref{fig:enter-label}).

\begin{figure*}[t!]
\centering
\includegraphics[width=1.0\textwidth]{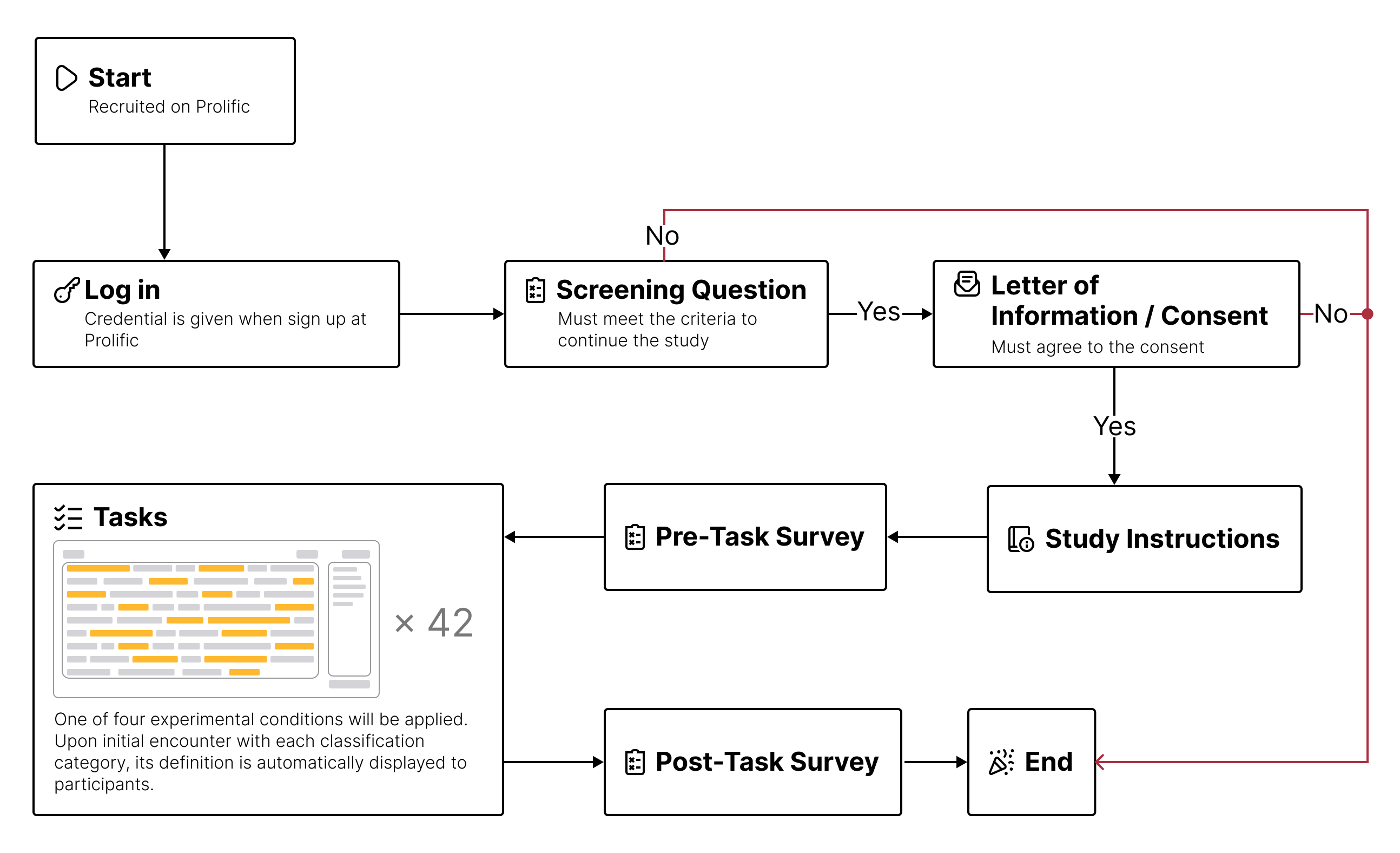}
\caption{Overview of study procedure participants followed in our study}
\label{fig:studyflow}
\end{figure*}

\subsection{Study Design}

Participants were randomly assigned to one of four (4) conditions and completed a pre-task survey rating their familiarity with annotation and medical tasks on a 7-point Likert scale. 
They were then presented with definitions and examples of all six (6) clinical categories.
After reading the definitions, they rated their perceived competence on a 5-point Likert scale (1 = not at all, 5 = extremely) based on two criteria — their \textit{understanding of each medical category} and their \textit{confidence in completing the annotation task for that category}.

In the main task, participants identified and highlighted words and phrases corresponding to a presented clinical category in clinical discharge summary documents, with one document and one category presented at a time. 
In AI conditions, recommendations appeared as highlighted text, alongside explanations (in the explanation condition). 
We randomly selected 20 discharge summaries from the MACCROBAT 2020 dataset \cite{Caufield2020}, out of which 7 random documents were selected for each participant, yielding up to 42 possible tasks per participant ($7\times 6 = 42$).
A document size larger than this resulted in increased participant fatigue, as observed during pilot studies.
The underlying NER model achieves 81.01\% precision, 80.04\% F1, and 79.10\% recall on this dataset \cite{pinard2025precision, raza2022large}. 
Participants were allocated an initial 20 minutes per session, with the option to continue or conclude afterward.
Upon finishing, participants completed a post-task survey on demographics, task experience, and re-rated their understanding of the clinical categories and confidence in their ability on a 5-point Likert scale.

\subsection{Participants}
We recruited 121 participants (71 female, 45 male, 3 non-binary, 2 undisclosed) via Prolific, requiring a minimum age of 18 and self-reported high English fluency. Participants were compensated at £9/hour (approx. 11.9 USD), with bonus pay at the same rate for time beyond 20 minutes.
Across the 4 experimental conditions, they completed an average of 13.84 tasks ($Min=1.00, Max=102.00, SD=13.93$), on an average of 3 documents and 6 categories for each document ($Min=1.00, Max=20.00, SD=2.99$).
We excluded 11 participants from the analysis who either did not complete the task, or performed only one action on a single task instance (refer to Table~\ref{tab:participants}).

\newcolumntype{R}{>{\raggedright\arraybackslash}X}  

\begin{table*}[t]
\centering

\begin{subtable}[t]{0.48\textwidth}
\centering
\caption{Sample composition}
\label{tab:participants-a}
\begin{tabularx}{\linewidth}{@{}lR@{}}
\toprule
\textbf{Characteristic} & \textbf{Value} \\
\midrule
\multicolumn{2}{@{}l}{\emph{Gender} (of 121 recruited)} \\
Female              & 71 \\
Male                & 45 \\
Non-binary          & 3 \\
Undisclosed         & 2 \\
\midrule
\multicolumn{2}{@{}l}{\emph{Condition} ($N$, analyzed)} \\
No AI               & 31 \\
No-AI + MD          & 28 \\
AI                  & 23 \\
AI + Exp            & 28 \\
\bottomrule
\end{tabularx}
\end{subtable}
\hfill
\begin{subtable}[t]{0.48\textwidth}
\centering
\caption{Engagement and pre-survey}
\label{tab:participants-b}
\begin{tabularx}{\linewidth}{@{}lR@{}}
\toprule
\textbf{Characteristic} & \textbf{Value} \\
\midrule
\multicolumn{2}{@{}l}{} \\
Tasks completed     & 13.84 (SD 13.93) \\
Documents           & 3 (SD 2.63) \\
Categories          & 6 (mean 4.74 engaged) \\
\midrule
\multicolumn{2}{@{}l}{\emph{Pre-survey self-report} ($M$, $SD$)} \\
Prior annotation experience (1-7) & 2.36 (1.77) \\
Healthcare domain knowledge (1-7) & 1.80 (1.40) \\
Task understanding  (1-5) & 4.42 (0.76) \\
Task confidence (1-5)  & 4.23 (0.89) \\
\bottomrule
\end{tabularx}
\end{subtable}
\vspace{0.3cm}
\caption{Sample characteristics. Gender counts are over all 121 recruited
participants; condition sample sizes and pre-survey statistics are for the final
analyzed sample ($N=110$).}
\label{tab:participants}
\end{table*}

\subsection{Measures}
\label{sec:measures}

To measure reliance calibration, we draw on the Brier Score framework \cite{glenn1950verification}, which computes calibration as the mean squared difference between predicted probabilities and observed outcomes. 
We adapt this framework to the reliance setting by treating the \textit{per-item reliance} indicator ($r_{ij}$) as the predicted probability and AI correctness as the observed binary outcome ($c_{ij}$), thereby preserving the scoring logic.
In this setting, lower RCS values indicate better calibration suggesting closer alignment between reliance behavior and actual AI correctness. 
We further decompose RCS into its over- and under-reliance components using Equations~\ref{eq:or} and~\ref{eq:ur}.
We compute the task-level Reliance Calibration Score (RCS) by aggregating the per-item reliance indicator (Equation~\ref{eq:reliance-indicator}) and corresponding AI correctness over each task (Equation~\ref{eq:rc}), and measured perceived competence through aggregated self-reported measures.


\subsubsection{Per-item reliance indicator}

For every AI-suggested annotation $j$ presented to participant $i$, we computed a per-item reliance indicator ($r_{ij}$) and AI correctness ($c_{ij}$). 
The interface logged three types of user interactions: (1) accepting the suggestion as is (taking no action on the highlighted span), (2) modifying it (editing its boundaries or label), or (3) deleting it. 
We map these actions onto a per-item reliance indicator ($r_{ij}$) as:

\begin{equation}
r_{ij} =
\begin{cases}
1 & \text{if } a_{ij} = \text{accept}, \\
0.5  & \text{if } a_{ij} = \text{modify}, \\
0 & \text{if } a_{ij} = \text{delete},
\end{cases}
\label{eq:reliance-indicator}
\end{equation}

$r_{ij} = 1$ reflects complete reliance on the AI suggestion, $r_{ij} = 0$ reflects complete rejection, and $r_{ij} = 0.5$ reflects partial reliance where the participant retained the suggestion but overrode part of its content.

$c_{ij} \in \{0, 1\}$ is a binary indicator of whether AI suggestion $j$ is correct, as determined by matching against the ground-truth annotations ($1$ if correct, $0$ if incorrect). 
Using established span-matching conventions from NER evaluation \cite{jiang2016evaluating}, we measured AI suggestion correctness across three criteria, exact boundary alignment, Intersection over Union (IoU) for span indices (threshold $\geq 0.5$), and token-level Jaccard similarity (threshold$\geq$ 0.8), and computed the overall AI correctness rate of 58.79\% on the given tested set of documents across the 6 clinical categories.

\subsubsection{Reliance Calibration Score (RCS)}
We computed reliance calibration as the degree to which a participant's reliance aligns with AI correctness across predicted items.
A well-calibrated participant accepts an AI suggestion when it is correct and rejects it when it is not.
We define Reliance Calibration score for a task as the mean squared difference between reliance and AI correctness over the $N$ AI-suggested items in that task. 

 \begin{equation}
\mathit{RCS} = \frac{1}{N} \sum_{j=1}^{N} \left( r_{ij} - c_{ij} \right)^2 .
\label{eq:rc}
\end{equation}

The Reliance Calibration Score, $\mathit{RCS}$ is an error score bounded in $[0, 1]$ \cite{glenn1950verification}.
Low values indicate that reliance closely aligns with AI correctness  suggesting good calibration, whereas high values indicate miscalibration in either direction. 
We compute $\mathit{RCS}$ at the task level per participant, where a task consists of annotating one document for one category.

\subsubsection{Over- and Under-reliance Decomposition}
 
Alongside $\mathit{RCS}$, we computed two separate linear measures that decompose calibration by direction. 
Over-reliance captures the extent to which a participant relied on \emph{incorrect} AI suggestions, including completely accepted or edited suggestions that did not match the ground truth. 
Under-reliance captures the extent to which a participant rejected \emph{correct} AI suggestions, including edited and deleted suggestions that did match the ground truth.
 
\begin{align}
\mathit{OR} &= \frac{1}{N} \sum_{j=1}^{N} r_{ij} \left(1 - c_{ij}\right), \label{eq:or} \\[4pt]
\mathit{UR} &= \frac{1}{N} \sum_{j=1}^{N} \left(1 - r_{ij}\right) c_{ij}.  \label{eq:ur}
\end{align}
 
\noindent Both scores are bounded in $[0, 1]$ and computed at the task level, using the same $r_{ij}$ and $c_{ij}$ as $\mathit{RCS}$ (Equation~\ref{eq:rc}). 
$\mathit{OR}$ is high when a participant frequently accepts AI suggestions that are incorrect, while $\mathit{UR}$ is high when a participant frequently rejects or edits suggestions that are correct. 
Because $r_{ij}$ is linear, a modified suggestion ($r_{ij} = 0.5$) contributes $0.5(1-c_{ij})$ to $\mathit{OR}$ and $0.5\, c_{ij}$ to $\mathit{UR}$, reflecting partial calibration rather than a binary outcome. 
These two measures inform the direction of miscalibration for each participant.

\subsubsection{Perceived Competence (PC)}

Perceived competence (PC) is a meta-cognitive construct reflecting individuals' self-assessment 
of their own abilities \cite{flavell1979metacognition, dunning2011dunning}. 
We operationalized it using pre- and post-task surveys in which participants rated (i) their understanding of each 
clinical category and (ii) their confidence in annotating it, after reading the corresponding definitions and examples. 
Pre-task perceived competence ($\mathit{PC}_{\mathit{pre}}$) and post-task perceived competence ($\mathit{PC}_{\mathit{post}}$) were each computed as the mean of understanding and confidence ratings across the six categories. 
The shift $\Delta\mathit{PC} = \mathit{PC}_{\mathit{post}} - \mathit{PC}_{\mathit{pre}}$ captures how self-assessed competence changed following task completion. 
These measures are used to address RQ3 and RQ4 (Sections~\ref{sec:rq3} and~\ref{sec:rq4}).

\subsection{Data Analysis}
We analyzed reliance calibration using Bayesian mixed-effects regression, estimated with the \textit{brms} package in R~\cite{carpenter2017stan, burkner2017brms}. 
We specified models we fit in each section. 
By default, each model included a by-participant random intercept to account for repeated tasks within annotators; models testing change across the session additionally included a by-participant random slope for task order, allowing individual trajectories to vary. 
For each model we ran four sampling chains for 2,000 iterations with a 1,000-iteration warm-up, yielding 4,000 posterior draws per parameter. 
We specified weakly informative priors: a Student's $t$-distribution ($\nu = 3$, $\mu = 0$, $\sigma = 2.5$) for the intercept, a Student's $t$-distribution ($\nu = 3$, $\mu = 0$, $\sigma = 2.5$) for the residual and random-effect standard deviations, a normal prior ($\mu = 0$, $\sigma = 10$) for the regression coefficients, and an LKJ(2) prior for the correlation between random intercepts and slopes.

We report posterior means and 95\% credible intervals (CIs) for each parameter, and consider an effect credible when its 95\% CI excludes zero. For directional hypotheses, we additionally report the posterior probability that the effect lies in the predicted direction ($P(\beta > 0)$ or $P(\beta < 0)$). 
For hypotheses of no effect, we assess practical equivalence by reporting the proportion of the posterior falling within a region of practical equivalence (ROPE) of $\pm$0.05 RC units. 

\begin{figure*}[t!]
\centering
  \includegraphics[width=\textwidth]{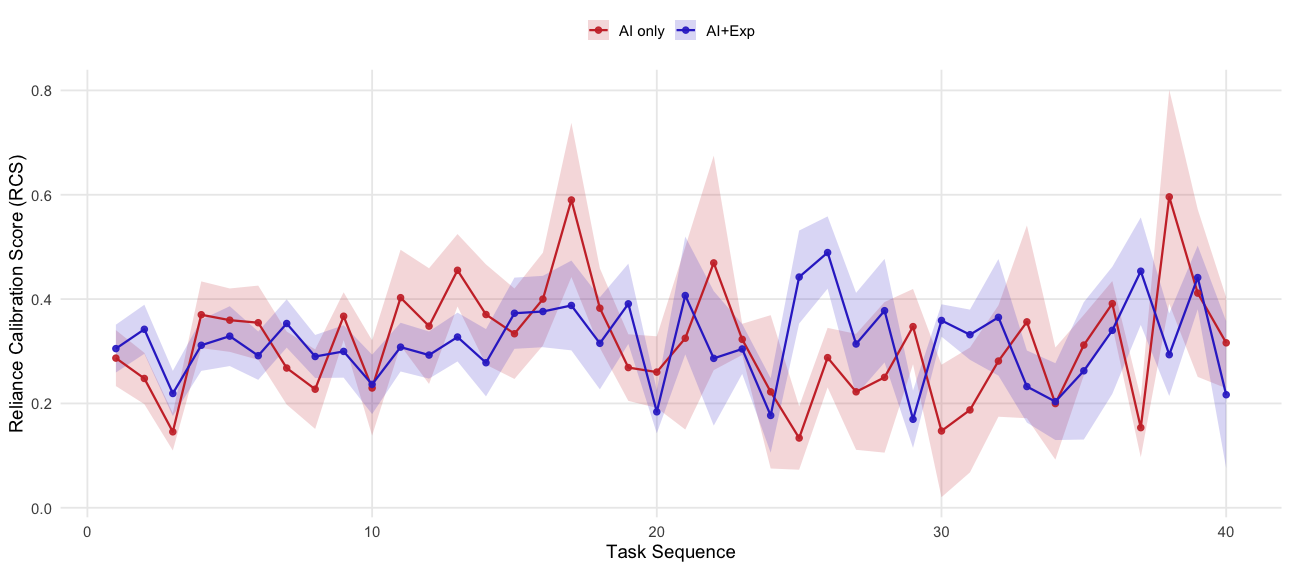}
  \caption{Comparison of changes in $RCS$ (lower = better
calibrated) across sequential tasks for each AI condition (AI and AI + Exp), presenting average calibration scores with standard errors. Single outlier data from the 41st task sequence was excluded.}
  \label{fig:rc_time}
\end{figure*}

\section{Results}

Across 4 experimental conditions, 110 participants provided 26,460 decision points in the form of annotated text.
The overall task accuracy varied across experimental conditions;
participants in AI-assisted conditions achieved higher total accuracy (\textit{AI}: $M = 0.429$, $SD = 0.257$; \textit{AI + Exp}: $M = 0.454$, $SD = 0.260$) compared to non-AI conditions (\textit{No AI}: $M = 0.211$, $SD = 0.226$; \textit{No AI + MD}: $M = 0.180$, $SD = 0.212$).

\subsection{RQ1: Session Trajectory of Reliance Calibration}
\label{sec:rq1}

\begin{figure*}[t!]
\centering
  \includegraphics[width=\textwidth]{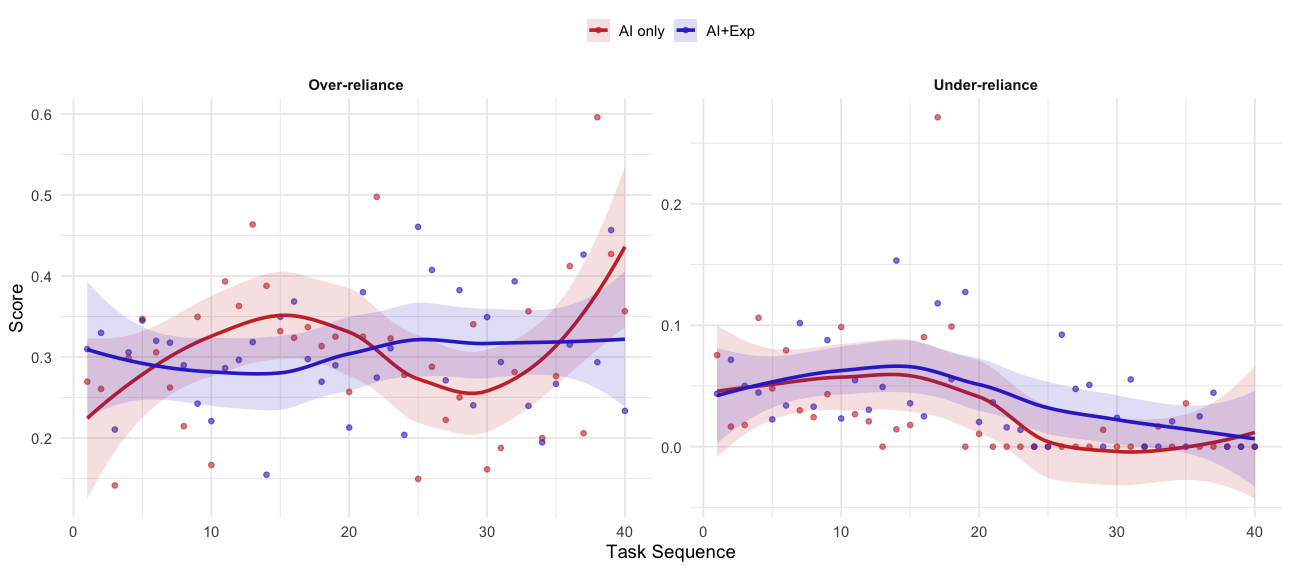}
  \caption{Comparison of changes in $OR$ and $UR$ across sequential tasks for each AI condition (AI and AI + Exp), presenting average calibration scores with standard errors. Lower OR and higher UR denote better-calibrated over- and under- reliance. Single outlier data from the 41st task sequence was excluded.}
  \label{fig:rc_over_under_time}
\end{figure*}

We compute the reliance indicator ($r_{ij}$; Equation~\ref{eq:reliance-indicator}) across all tasks in the \textit{AI} and \textit{AI + Exp} conditions, yielding a mean of $0.85$ ($SD = 0.28$). 
Participants accepted AI suggestions $75.0$\% of the time, modified them $20.2$\% of the time, and rejected them $4.8$\% of the time across these two conditions.
Acceptance was lower when explanations accompanied predictions (\textit{AI + Exp}: $72.0$\%) compared to predictions alone (\textit{AI}: $79.0$\%). 
We observed a mean RCS of $0.324$ ($SD = 0.247$), with over-reliance substantially higher than under-reliance (OR: $M = 0.31$, $SD = 0.24$; UR: $M = 0.04$, $SD = 0.12$).

To examine whether miscalibration changed across the session, we fit a Bayesian linear mixed-effects model regressing RCS on task sequence with random intercepts and slopes for participants. 
We observed that the RCS tended to increase over the session, though the effect was not significant.
Over-reliance increased as the session 
progressed (Estimate $= 0.04$, $95\%$ CI $[-0.01, 0.09]$; $P(\beta > 0) = 0.94$), while under-reliance decreased significantly (Estimate $= -0.01$, $95\%$ CI $[-0.03, 0.00]$; $P(\beta < 0) = 0.97$), suggesting a drift towards uncritical acceptance of AI suggestions over time. 
Random-slope variance exceeded the average slope, indicating substantial individual variation in this trajectory. 
A paired one-sided Wilcoxon signed-rank test confirmed that participants over-relied 
significantly more than they under-relied ($V = 1205$, $p \ll .001$).

We further fit a series of Bayesian mixed-effects models on RCS. 
The intercept-only model (\textbf{M0}) estimated a grand mean RCS of $0.32$ ($95\%$ CI $[0.30, 0.34]$) with negligible 
between-participant variance ($SD = 0.02$; $ICC \approx 0.006$). 
Adding the clinical category as a fixed 
effect (\textbf{M1}) revealed substantial variation across annotation types where tasks related to the \textit{Lab Value} category produced the highest miscalibration (estimated mean $\approx 0.60$, $\beta = 0.33$, $95\%$ CI $[0.28, 0.38]$), while Diagnostic Procedure showed moderate miscalibration ($\beta = 0.07$, $95\%$ CI $[0.02, 0.13]$). 
Category reduced residual variance from $0.25$ to $0.21$. 
Adding document ID (\textbf{M3}) as a predictor yielded no evidence ($\beta \approx 0.00$, $95\%$ CI $[0.00, 0.00]$), with category coefficients and residual variance unchanged.

\subsection{RQ2: Effect of AI Explanations on Reliance Calibration}
\label{sec:rq2}


We examined whether AI explanations affected reliance calibration (\textbf{H2}) using a Bayesian linear mixed-effects model with participant random intercepts and explanation condition and clinical category as fixed effects. 
The explanation coefficient was negligible and non-credible ($\beta = -0.02$, $95\%$ CI $[-0.05, 0.02]$), with the posterior centered near zero and residual variance unchanged from M1 ($\sigma = 0.21$). 
The category-level miscalibration pattern was similar to M1, indicating that explanation condition did not attenuate difficulty differences across clinical category types.

\begin{figure*}[t!]
\centering
  \includegraphics[width=0.7\textwidth]{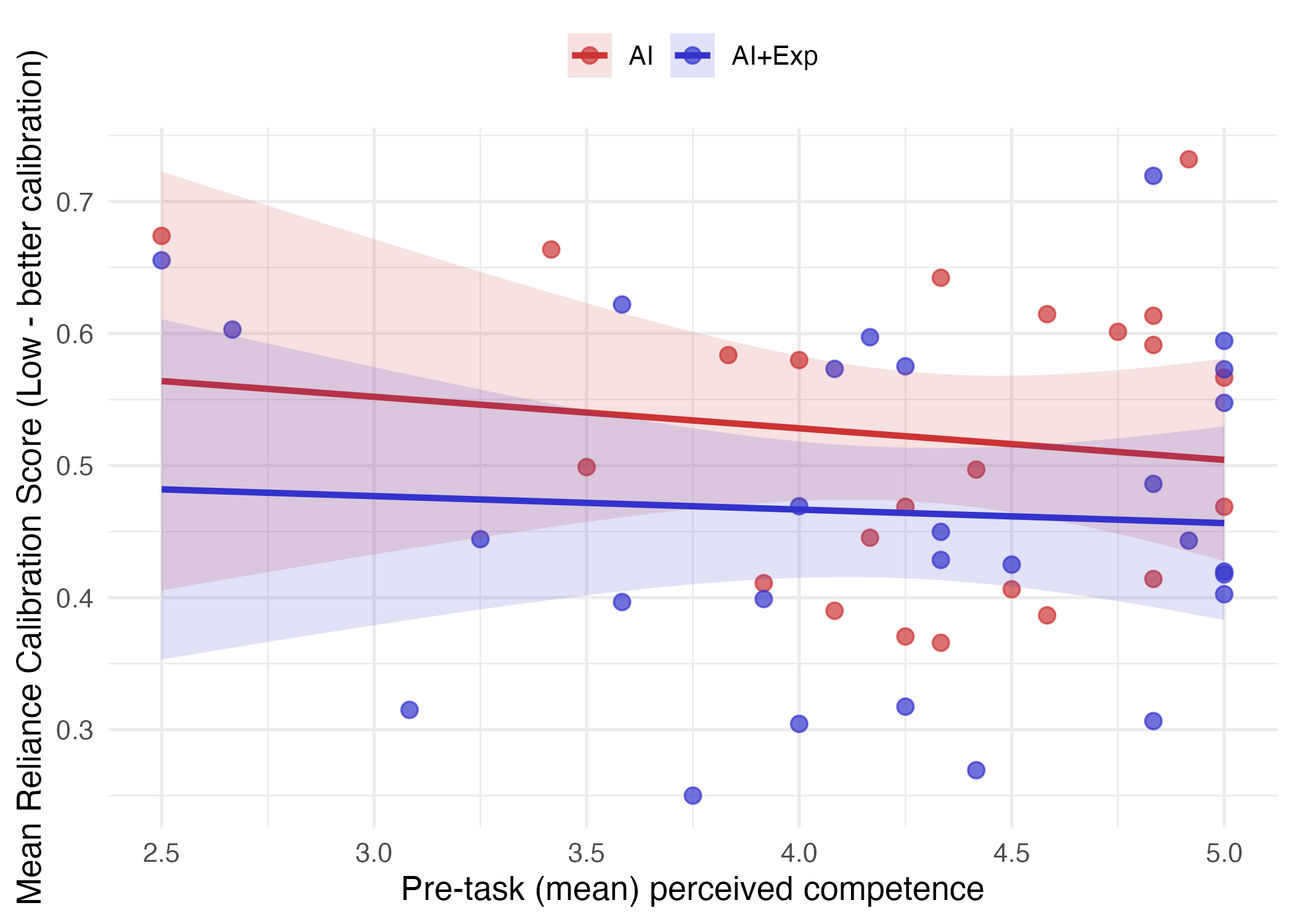}
  \caption{Per-participant mean RCS (lower = better
calibrated) against pre-task perceived competence (1--5).}
  \label{fig:rq3}
\end{figure*}

Within the \textit{AI + Exp} condition, we tested whether active engagement with the confidence score and medical dictionary, measured by dwell time and number of clicks to access these, predicted better reliance calibration. 
Neither confidence score dwell time ($\beta = 0.00$, $95\%$ CI $[-0.01, 0.02]$) nor dictionary dwell time ($\beta = -0.01$, $95\%$ CI $[-0.03, 0.01]$) was associated with lower RCS.
We further modeled over- and under-reliance separately, with clinical category as a fixed effect and by-participant random intercepts. 
While explanations had no significant effect on over-reliance or under-reliance,
the per-item reliance indicator ($r_{ij}$) was positively associated with AI correctness in both conditions.
The relationship was weak in \textit{AI Only} ($\rho = 0.20$, $p < .001$) and slightly stronger in \textit{AI + Exp} ($\rho = 0.27$, $p < .001$). 
A Fisher $r$-to-$z$ test confirmed these correlations differed significantly ($z = -2.52$, $p = .012$), suggesting that explanations modestly improved annotators' sensitivity to AI correctness.

\subsection{RQ3: Perceived Competence and Reliance Calibration}
\label{sec:rq3}

We observed that the pre-session perceived competence was uniformly high across all conditions (PC$_\text{pre}$: $M = 4.33$, $SD = 0.61$) and declined slightly by session end (PC$_\text{post}$: $M = 4.03$, $SD = 0.86$).
The drop was largest in the \textit{AI} condition ($M = -0.47$, $SD = 0.88$), compared to \textit{AI + Exp} ($M = -0.21$, $SD = 0.59$), \textit{No AI} ($M = -0.26$, $SD = 0.61$), and \textit{No AI + MD} ($M = -0.30$, $SD = 0.73$). 
A Bayesian mixed-effects model with explanation and PC$_\text{pre}$ as fixed effects and participant as a random intercept showed a small negative point estimate for PC$_\text{pre}$ ($\beta = -0.03$, $95\%$ CI $[-0.08, 0.02]$), with the interval crossing zero. 
We used PC$_\text{pre}$ as the moderator to ensure temporal precedence over the experimental manipulation, given that post-session scores were likely shaped by explanation exposure.
A directional hypothesis test provided moderate support for \textbf{H3a} where the posterior probability that higher perceived competence predicted lower RCS was $0.88$ ($ER = 7.08$), suggesting a negative relationship between PC$_\text{pre}$ and reliance miscalibration.
The interaction model revealed a credible negative main effect of PC$_\text{pre}$ on RCS ($\beta = -0.03$, $95\%$ CI $[-0.14, -0.08]$), suggesting that participants with higher pre-session perceived competence showed lower miscalibration in the \textit{AI} condition (\textbf{H3a}).
The explanation $\times$ PC$_\text{pre}$ interaction was positive but non-credible ($\beta = 0.05$, $95\%$ CI $[-0.02, 0.13]$), with posterior mass predominantly in the positive direction, suggesting that contrary to \textbf{H3b}, explanations may have narrowed rather than amplified the calibration advantage of high-competence participants.
Finally, PC$_\text{pre}$ showed a consistent directional effect on over-reliance ($\beta = -0.03$, $95\%$ CI $[-0.07, 0.01]$; $P(\beta < 0) = 0.91$; $ER = 9.55$), while the under-reliance model yielded null effects for both PC$_\text{pre}$ ($\beta = 0.00$, $95\%$ CI $[-0.01, 0.02]$) and explanation condition ($\beta = 0.00$, $95\%$ CI $[-0.02, 0.02]$).


\begin{figure*}[t!]
\centering
  \includegraphics[width=1\textwidth]{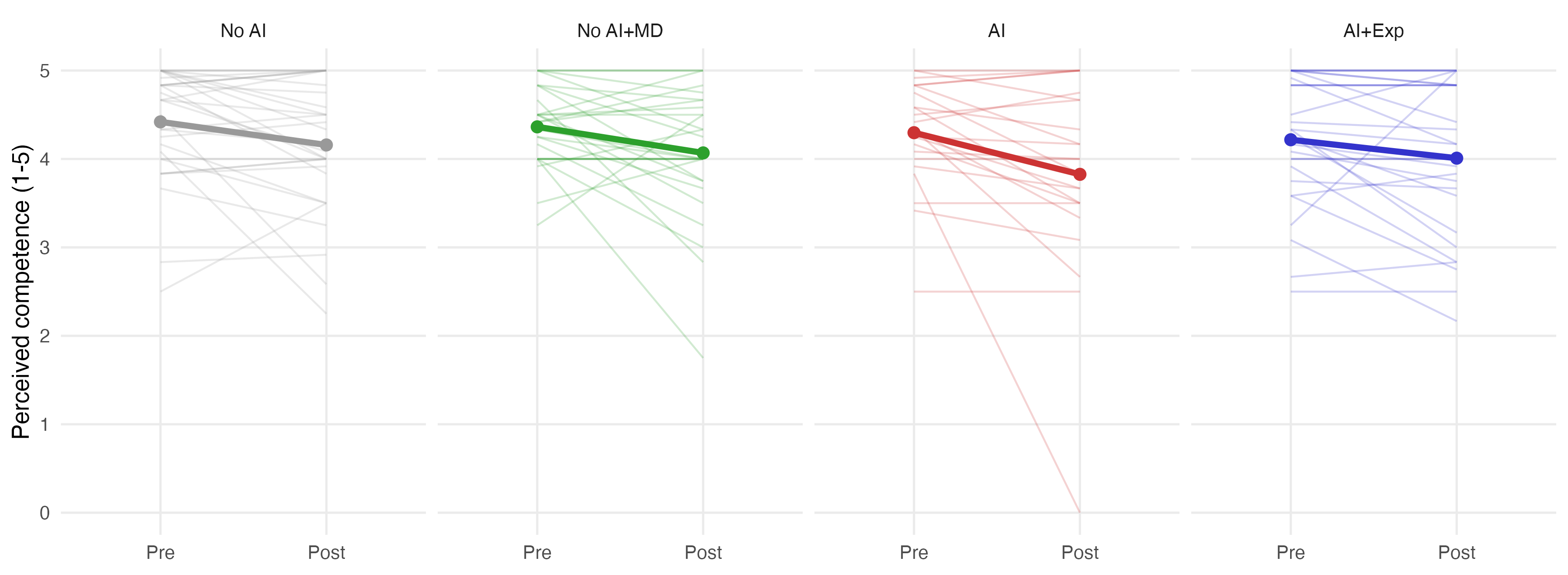}
  \caption{Participant pre$\rightarrow$post change in mean perceived competence by condition. Thin lines are
individual annotators. Thick lines are condition means.}
  \label{fig:rq3_2}
\end{figure*}

\subsection{RQ4: Confidence vs. Understanding and Reliance Calibration}
\label{sec:rq4}

We tested whether self-reported confidence in task ability predicted better-calibrated reliance independently of task understanding.
For (\textbf{H4}), we operationalized better calibration as a stronger positive coupling between AI suggestion correctness and user actions (accept, reject, and edit). 
A well-calibrated annotator accepts the AI more readily when it is correct than when it is wrong,
as well as rejects or modifies the AI more readily when it is incorrect.
Pre-session confidence and understanding were high in both AI conditions and declined modestly after the task (\textit{AI}: Confidence$_\text{pre}$ $= 4.21$, Understanding$_\text{pre}$ $= 4.38$; \textit{AI + Exp}: Confidence$_\text{pre}$ $= 4.11$, Understanding$_\text{pre}$ $= 4.32$). 
We estimated separate confidence and understanding slopes for each action channel in a Bayesian logistic mixed model ($N = 12{,}798$ items).
While Confidence$_\text{pre}$ did not predict appropriate reliance in either acceptance channel effects on accepting correct suggestions, there was a small effect on adding missed entities ($\beta = 0.03$, $95\%$ CI $[-0.36, 0.42]$). 
For incorrect suggestions, more confident participants were less likely to reject or edit wrong suggestions ($\beta = -0.33$, $95\%$ CI $[-0.73, 0.05]$; $P(\beta < 0) = 0.95$), suggesting that confidence may increase over-reliance on incorrect AI output.

Understanding$_\text{pre}$ behaved oppositely, in the sense that higher understanding increased both rejection and editing of incorrect suggestions ($\beta = 0.51$, $95\%$ CI $[0.09, 0.94]$; $P(\beta > 0) = 0.99$) and addition of missed entities ($\beta = 0.47$, $95\%$ CI $[0.05, 0.89]$; $P(\beta > 0) = 0.99$), while showing no credible effect on accepting correct suggestions ($\beta = -0.31$, $95\%$ CI $[-0.75, 0.14]$; $P(\beta > 0) = 0.09$). 
These results suggest that within perceived competence, understanding drives appropriate rejection of incorrect AI suggestions, while confidence alone does not improve calibration and may marginally increase over-reliance.

\begin{figure*}[t!]
\centering
  \includegraphics[width=0.8\textwidth]{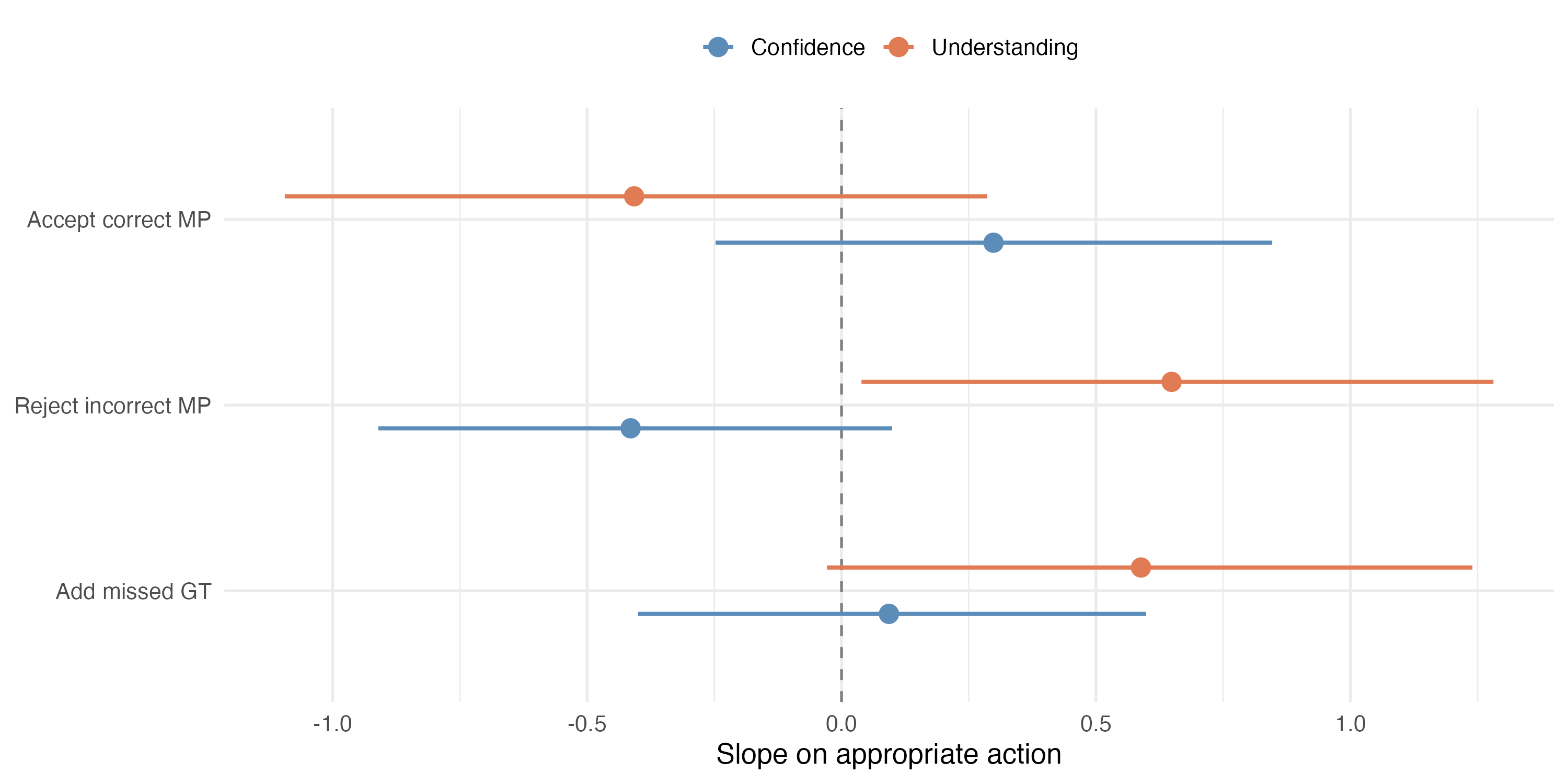}
  \caption{Task Confidence (Pre) vs Task Understanding, per appropriate reliance behavior}
  \label{fig:rq4}
\end{figure*}

\section{Discussion}

Our findings reveal that reliance calibration in feedback-free environments is shaped by the interplay of task knowledge structure, explanation design, and users' meta-cognitive beliefs. 
In this section, we discuss the implications of these findings for the design of human-AI collaborative systems.

\subsection{Designing Meta-Cognitive Functions for Sustaining Appropriate Reliance}

We observed that reliance miscalibration varied substantially across clinical categories and the same user working with the same AI system, exhibited different reliance behavior depending on the task.
For instance, task category of \textit{Lab Value} produced significantly higher miscalibration than all other categories.
This was likely because numeric values are visually salient and easy to identify, leading users to over-rely on AI suggestions without evaluating whether the AI recommendation meets the clinical definition of a lab value in context \cite{cummings2017automation}.
One of the participants in \textit{AI} confirmed this rationale (P81: \textit{`For tasks that were easy to understand (medication and lab value), I simply skimmed through the text to find the appropriate text without taking the AI highlighted text into account'}. 
These observations suggest that reliance calibration shifts in response to the knowledge demands of specific task types.
Instead of studying reliance as a global user characteristic, designers of human-AI collaborative systems may monitor knowledge-specific reliance patterns.

One way to do this is to embed \textit{meta-cognitive monitoring functions} in AI-assisted tools that model knowledge-specific reliance patterns and use them to scaffold user reasoning. 
Unlike performance feedback, which requires ground truth to be available during the task, such functions can operate on behavioral signals by monitoring shifts in acceptance rates, editing frequency, and dwell time on explanations. 
For instance, a spike in acceptance rate within a knowledge domain, or a collapse in post-recommendation editing behavior relative to a user's interaction baseline, could trigger appropriate interventions.
These interventions could include reflection prompts that draw the user's attention to their reliance patterns, or a nudge to use explanations as validation tools. 
This approach is particularly valuable in human-AI collaborative environments that do not provide immediate feedback on user performance. 
Future explainable AI systems could thus shift from \textit{outcome-oriented transparency} towards \textit{process-oriented meta-cognitive support}, treating knowledge-specific behavioral patterns as real-time signals to be monitored and scaffolded.

\subsection{Alternate role of Explanations as Meta-Cognitive Scaffolds}

Our results showed that neither the presence of AI explanations nor active engagement with them improved reliance calibration, contradicting prior research in which explanation support reliance behavior in feedback-rich settings \cite{schemmer_appropriate_2023, vasconcelos2023explanations, buccinca2021trust}. 
This failure reflects a mismatch between how explanations are designed and the role they are expected to play.
Presented uniformly with every suggestion, they place the burden of evaluation on users, 
compounding cognitive fatigue across sequential tasks, especially for those lacking the epistemic resources to use them accurately. 
Our finding that perceived understanding, and not confidence, predicted appropriate rejection of incorrect suggestions points to an alternative role for explanations in improving domain comprehension and critical evaluation. 

Explanations can serve as \textit{meta-cognitive scaffolds} that support users in monitoring and regulating their own reasoning rather than verifying AI output. 
This also means, that designers of human-AI collaborative systems must also rethink how and when explanations are delivered.
One strategy is to deliver explanations when a user's trust level changes \cite{srinivasan2026adjust}.
While helpful, it may not always work, especially in settings where the AI is more accurate than the user and high trust is warranted.
Another strategy is to deliver explanations when a \textit{calibration discrepancy} occurs, where a user's reliance behavior in a specific knowledge domain diverges from the AI's estimated correctness in that domain.
For instance, if a user consistently accepts AI suggestions for Lab Value annotations at a rate of 90\% across a session, while the AI's precision for that category is 65\%, the system can infer a likely calibration discrepancy.
In feedback-free environments, such explanations could serve as a proxy for the corrective signal that performance feedback would otherwise provide \cite{buccinca2026offline, gajos2022people}.

\subsection{Capturing User Rationale to Support Meta-Cognitive Monitoring}

The participants developed informal strategies for self-verification.
For instance, in post-task reflection, participants stated that they tried to recognize emerging patterns from the content in task instances (P3 (No AI):`\textit{I looked for patterns as I moved on to new texts, and [...] compare the ways in which the texts were formatted or structured similarly [...]}', P51 (AI): \textit{`Most tasks end with a y, most medications are compound words, most values are numerical'}). 
These behaviors represent spontaneous meta-cognitive monitoring in the absence of external feedback which modern AI systems fail to capture or reinforce. 
Designers of future human-AI collaborative systems should treat user rationale as a resource for supporting appropriate reliance calibration. 
For instance, asking users to tag whether an acceptance was pattern-based or definition-grounded could help identify superficial reasoning before it consolidates into systematic over-reliance. 
These micro-rationales build a behavioral profile of why a user relies as they do, not just how much. 
At natural breakpoints (e.g., after completing all categories for a document), the system could show users a brief summary of their decision rationales alongside their pre-category acceptance rates.
This can create moments of reflection that approximate performance feedback without requiring ground truth.

\section{Limitations}
We acknowledge several limitations regarding our task, participant population and prototype.
First, we selected the task of annotating clinical concepts, because they are easily distinguishable and contain expressive knowledge across categories. 
If the categories are indistinguishable and difficult to identify, even with instructions and AI explanations, misunderstandings on the information may bias the results.
Secondly, we recruited participants via Prolific, a research-purposed crowd-sourcing platform, which does not represent a wide demographic.
Finally, while we provided motivation in terms of monetary reward, we did not account for variables such as cognitive load and implicit motivation that play a key role in system usage.

\section{Conclusion}
We investigated how novice users calibrate their reliance on AI in a clinical entity annotation task when no external feedback is available, and how their meta-cognitive beliefs shape this process. We found that reliance calibration in this setting was neither improved by AI explanations nor reliably guided by users' confidence in their own competence; instead, it depended on their genuine understanding of the task domain and varied with the kind of knowledge involved. These findings suggest that supporting appropriate reliance for novices without feedback calls for scaffolding meta-cognitive engagement rather than surfacing AI confidence, and motivate further study of reliance calibration in the many real-world settings where feedback is absent.


\bibliographystyle{ACM-Reference-Format}
\bibliography{sample-base}

\appendix

\end{document}